\documentclass[aps,showpacs,graphicx]{revtex4} 
\usepackage[dvips]{graphicx} 
\usepackage[english]{babel} 
\usepackage{rotating}
\usepackage{lscape}
\newcommand{\beq}{\begin{equation}} 
\newcommand{\eeq}{\end{equation}} 
\newcommand{\beqn}{\begin{eqnarray}} 
\newcommand{\eeqn}{\end{eqnarray}}  
\newcommand{\nue}{\nu_e}
\newcommand{\num}{\nu_{\mu}}

\newcommand{\nux}{\nu_x}
\newcommand{\anue}{\overline{\nu}_e}
\newcommand{\anum}{\overline{\nu}_{\mu}}

\newcommand{\bqu}{{\bf q}}

\newcommand{\bk}{{\bf k}}
\newcommand{\epsn}{\epsilon_\nu}
\newcommand{\trec}{T_{\rm rec}}
\newcommand{\etre}{E_{\rm th}}
\newcommand{\fsns}{f^{\rm SNS}}
\newcommand{\rc}{{\rm RC}\,} %Reference calculation
\newcommand{\rr}{{\cal R}}            %radial wave function
\newcommand{\half}{\frac{1}{2}}

\usepackage[usenames]{color} 
\usepackage{ulem}

\begin{document} 

%\noindent 
\title{Nuclear structure uncertainties in coherent elastic neutrino-nucleus scattering}

\author{G. Co'$^{\,1,2}$, M. Anguiano$^{\,3}$, A. M. Lallena$^{\,3}$}
\affiliation{
$^1$ Dipartimento di Matematica e Fisica ``E. De Giorgi'', 
  Universit\`a del Salento, I-73100 Lecce, Italy \\ 
$^2$ INFN Sezione di Lecce, Via Arnesano, I-73100 Lecce, Italy 
$^3$ Departamento de F\'\i sica At\'omica, Molecular y 
  Nuclear, Universidad de Granada, E-18071 Granada, Spain \\
}  

\date{\today}

\bigskip 
 
\begin{abstract} The effects of the nuclear structure uncertainties on the description 
of processes induced by coherent scattering of neutrinos on 
nuclei are investigated. A reference calculation based on a specific nuclear 
model is defined and the cross sections 
and also the expected number of events
produced by neutrinos generated by the explosion of a supernova in our galaxy,
and by a spallation neutron source are evaluated. 
By changing the input parameters of 
the reference calculation their relevance on cross sections and on 
the number of the detected events is estimated. 
Seven spherical nuclei with different proton to neutron ratios
are considered as possible targets of the neutrinos in the detector, 
the lightest being $^{12}$C and the heaviest $^{208}$Pb. 
The effects generated by the uncertainties 
of the nuclear model are much smaller than those due to the supernova neutrino flux models.
This makes the coherent elastic neutrino-nucleus scattering a reliable tool
to investigate the details of the neutrino sources, the neutrino-nucleus interaction,
and, eventually, also to extract information about neutron distributions in nuclei.
\end{abstract} 


\maketitle 

%-------------------------------
\section{Introduction}
\label{sec:intro}

Elastic neutrino scattering processes where the energy of the
recoiling nucleus is detected, was proposed
some decades ago as an investigation
tool for a plethora of conventional neutrino physics topics
and new-physics open issues \cite{fre74,dru84}.  
Since for neutrinos with energy up to about 100 MeV 
the individual nucleonic scattering amplitudes
overlap in phase \cite{bed18}, 
the process is called \textsl{Coherent} Elastic Neutrino-Nucleus
Scattering (CE$\nu$NS) \cite{sch15}.  Due to its elastic character,
there is not a neutrino energy threshold preventing CE$\nu$NS  to
happen and, thus, for neutrino energies of few tens of MeV, CE$\nu$NS  cross
sections are remarkably larger than those of other competing processes 
\cite{aki17a,ric19,lin18}. 
The experimental study of CE$\nu$NS  shares various technical 
problems with direct dark matter searches, therefore many detectors
take advantage of the experiences accumulated in this latter investigation
field. 

The first observation of CE$\nu$NS  was reported by the COHERENT
collaboration \cite{aki17a}, making use of the pulsed neutrino
emission of the Spallation Neutron Source (SNS) at Oak Ridge National
Laboratory.  Following this pioneering experience, a 
remarkable number of new experiments has been planned: 
CONNIE \cite{agu16}, CONUS \cite{lin18},
MINER \cite{agn17}, $\nu$-gen \cite{bel15}, RED-100 \cite{aki17b},
RICOCHET \cite{bil17}, TEXONO \cite{won10} NU-CLEUS
\cite{str17a,rot19}.

From the theoretical point of view, the main uncertainties in the
evaluation of the CE$\nu$NS cross sections come from the proton and
neutron density distributions of the target nuclei. 
Proton density distributions can be obtained from
the experimental charge densities extracted from elastic
electron scattering data \cite{dej87}. Unfortunately, 
these empirical charge distributions are known only for a limited number of
 nuclei. Furthermore, one has to consider that
the main contribution to CE$\nu$NS  
cross sections is due to the neutron density distributions which are
hardly constrained by experimental data. 
As a consequence, the information about proton
and neutron density distributions 
is mainly based on nuclear models. In
these last years these models have improved a lot their performances in
describing nuclear ground states, and, even if formulated in different
manners, they show a high degree of agreement in their results
\cite{co12a}.

The present study aims at evaluating the effects 
of  the ambiguities inherent to the description 
of the proton and neutron distributions 
on the CE$\nu$NS cross sections. 
The strategy of our work consists in defining a reference calculation (RC)
where a specific nuclear model, which we have
developed, tested and perfected in these last years
\cite{ang14,ang15,ang16a,co18a,ang19}, is used, and then in comparing its results 
with those of calculations carried out by, reasonably, modifying the 
input parameters.

A set of nuclei of interest for CE$\nu$NS experiments, and 
selected in various regions of the nuclear chart, has been investigated.
We considered
$^{12}$C, $^{16}$O, $^{20}$Ne, $^{40}$Ar, 
$^{76}$Ge, $^{132}$Xe and $^{208}$Pb.
Since our model can be applied only to spherical nuclei, 
we neglected the slight deformation of the ground states of $^{12}$C 
and $^{20}$Ne nuclei.

In section \ref{sec:cenns}, the expressions used to calculate
the CE$\nu$NS cross section are shown and related to the proton and neutron
density distributions obtained with our model. The details of
the nuclear model used to perform the RC are presented
in section \ref{sec:nucmod}. Our model has been applied to two specific
situations generating neutrinos, and interesting for CE$\nu$NS detection:
the case of a supernova explosion in our
galaxy, and the neutrino production induced by a 
spallation neutron source (SNS) \cite{sch06}. 
The information regarding the input parameters needed to 
estimate the number of detected events expected for
these two physical cases is presented
in section \ref{sec:super}.

In section \ref{sec:results}, we study how the CE$\nu$NS
cross sections and the number of detected events are modified 
under various hypotheses about the shape of nuclear form factors.
These changes are compared to those induced by the astrophysics uncertainties.

In section \ref{sec:conclusions} we summarize the results of our
study and conclude that, from the nuclear physics point of
view, the CE$\nu$NS processes are well understood.
This makes CE$\nu$NS a reliable tool
to investigate sources and properties of supernova neutrino fluxes.

%--------------------------------------
\section{The model}

\subsection{The CE$\nu$NS cross section}
\label{sec:cenns}
In this section, we present the basic equations
describing the elastic scattering of neutrinos off
a nucleus. In our expressions we shall use
natural units ($\hbar = c =1$).
The elastic neutrino scattering process is ruled by weak neutral currents,
and we describe them
by considering that a single $Z^0$ boson 
is exchanged between the
neutrino and the target nucleus. 
For elastic scattering, in the center of mass 
reference system, the relation between 
the neutrino initial and final momenta, $\bk_i$ and $\bk_f$ respectively,
and the neutrino energy $\epsn$  is 

\beq
|\bk_i|=|\bk_f|= \epsn \, .
\label{eq:que}
\eeq
In the laboratory frame, one has to take into
account the recoil of the target, which is zero in the limit of an
infinite target mass. 
Because the nuclear rest masses are much larger than the neutrino
energies, the relation (\ref{eq:que}) is numerically satisfied also in the 
laboratory frame, therefore
the momentum transfer $\bqu = \bk_i - \bk_f$ 
depends only on $\epsn$ and on the angle $\theta$
between $\bk_i$ and $\bk_f$, specifically,
\beq 
q^2 \,=\, ( \bk_i \,-\, \bk_f)^2 \, = \, 2 \,\epsn^2 \,(1 \,-\, \cos \theta) \, ,
\label{eq:q2}
\eeq 
where $q=|\bqu|$. The recoil energy of the target nucleus can be
expressed as 
\beq 
\trec \,=\, \frac {q^2}{2 \,M} \, ,
\eeq 
where $M$ indicates the rest mass of the nucleus.
For neutrino energies of the order of tens of MeV the nuclear
recoil energies are of the order of keV.

We evaluate the cross section by using traditional trace techniques
\cite{bjo64}, and we obtain, (see eq.~(9) of ref. \cite{pap15}),
\beq
\frac {{\rm d}\sigma} {{\rm d}(\cos \theta)} \,=\, 
\frac {G_{\rm F}^2} {2 \, \pi} \, \epsn^2 \, (1\, + \, \cos \theta) \, \left| {\cal A } \right|^2 \, ,
\label{eq:xscos}
\eeq
where $G_{\rm F}=1.1663787\cdot 10^{-11}\,{\rm MeV}^{-2}$ 
is the Fermi constant, and ${\cal A}$ the nuclear transition 
amplitude. In terms of the nuclear recoil energy $\trec$ we obtain
the expression, %(see eq.~(11) of  ref. \cite{pap15}),
\beq
\frac {{\rm d}\sigma} {{\rm d} \trec } \,=\, 
\frac{G^2_{\rm F} \, M}{\pi} \, 
\left(1\, -\,  \frac {M\, \trec} {2 \epsn^2} \right) \,
\left| {\cal A} \right|^2 \, .
\label{eq:xstrec}
\eeq

The nuclear transition amplitude can be expressed as \cite{pap15}
\beq
{\cal A} \,=\, 
\left( \half \,-\, 2 \, \sin^2 \theta_{\rm W} \right)\, Z\, F_{\rm p} (q) \, 
-\, \half \, N\,  F_{\rm n}(q) \, ,
\label{eq:mtrans}
\eeq
where $\theta_{\rm W}$ is the Weinberg angle, 
with $\sin^2 \theta_{\rm W}=0.23129$, $Z$ and $N$ are the proton and 
neutron numbers, respectively, and $F_{\rm p}$ and $F_{\rm n}$ 
are the proton and neutron nuclear form factors defined as
\beq
F_{\rm p}(q) \,=\,  \frac {4\, \pi}{Z} \displaystyle  
\int {\rm d}r \, r^2 \, \frac{\sin(qr)}{qr} \, \rho_{\rm p}^{\rm fold} (r) 
\label{eq:formf-p}
\eeq
and
\beq
F_{\rm n}(q) \,=\,  \frac {4\, \pi}{N} \displaystyle  
\int {\rm d}r \, r^2 \, \frac{\sin(qr)}{qr} \, \rho_{\rm n}^{\rm fold} (r) \, .
\label{eq:formf-n}
\eeq
The proton and neutron, folded, density distributions are normalised
to the proton and neutron numbers.
In this manner, the value of the 
nuclear form factors (\ref{eq:formf-p}) and (\ref{eq:formf-n})
in the limit for $q \rightarrow 0$ is 1. 

An appropriate description of the interaction 
between the Z$^0$ and each nucleon
requires to go beyond the approximation where
nucleons are assumed to be dimensionless and  
without internal structure. This is the reason why $F_{\rm p}$ and $F_{\rm n}$
are defined in terms of the corresponding folded density distributions:
\beq
 \rho_i^{\rm fold} (r) \,=\,
 \displaystyle  \int {\rm d}r' \, \rho_i(r\,-\,r') \, G^i_{\rm  w}(r') \, = \, 
 \frac {1}{2\, \pi} \,\int \, {\rm d}p \, \exp({i\,p\,r})\, \tilde{\rho}_i(p) \, 
 \tilde{G}^i_{\rm w}(p) \, , \,\,\, i\equiv {\rm p,n} \, .
 \label{eq:rhofold}
\eeq
Here $\rho_i(r)$ is the point-like density distribution and 
$G^i_{\rm w}$ the nucleon weak  form factor. Folded densities are actually
calculated as indicated in the second equality where $\tilde{\rho}_i(p)$ and 
$\tilde{G}^i_{\rm w}(p)$ are the Fourier transforms of the point-like density distribution 
and of the weak nucleon form factor, respectively. 

The isotopic invariance of the strong interactions allows us 
to connect the nucleonic weak form factor to the electromagnetic
one \cite{alb02, bot04t}, therefore
\beq
\tilde{G}^i_{\rm w}(p)\, = \, G_{\rm E}(p) \, ,
\label{eq:gnge}
\eeq
where $G_{\rm E}(p)$ is the Sachs proton electric form factor. 
In our calculations we have used the simple dipole parameterization:
\beq
G_{\rm E}(p) \,=\, \frac {1} {(1\,+\, 0.05833 \, p^2)^2} \, .
\label{eq:dipole}
\eeq

We obtain the total cross section by integrating eq.~(\ref{eq:xscos}) 
on $\cos \theta$ or eq.~(\ref{eq:xstrec}) on $\trec$:
\beq
\sigma  \, =\, 
\displaystyle \int_{-1}^1 {\rm d}(\cos \theta)\, \frac {{\rm d}\sigma} {{\rm d} (\cos \theta)} \, =\, 
\int_0^{\trec^{\rm max}} {\rm d}\trec \, \frac {{\rm d}\sigma} {{\rm d} \trec } \, .
\label{eq:xstot}
\eeq
For a given value of the neutrino energy, $\epsn$, the maximum value of 
the target recoil energy is
\beq
\trec^{\rm max} \,=\, \frac {2\, \epsn^2} {M}  \, .
\eeq

%--------------------------------------------------------------------------------------------------------
\subsection{The nuclear model of the reference calculation}
\label{sec:nucmod}

In the calculations of the CE$\nu$NS cross sections, 
all the information regarding
the nuclear target is contained
in the proton and neutron form factors that depend only on
the nucleon density distributions $\rho_i(r)$ of eq.~(\ref{eq:rhofold}). 
In our RC we use the nucleon distributions obtained with a Hartree-Fock (HF) 
plus Bardeen-Cooper-Schrieffer (BCS) model that we have developed in these
last years \cite{ang14,ang15,ang19}.

As already stated above, the nuclear model used in the present investigation is
only valid for spherical nuclei. The solution of
the corresponding spherical HF equations \cite{rin80,suh07} provides a set of
single-particle (s.p.) levels characterized by the orbital and total angular
momenta, $l$ and $j$ respectively, and the principal quantum 
number that we label as $n=1,2,\ldots$. The ground state of the nucleus 
considered is built up by filling separately the various proton and neutron s.p. levels,
in ascending order of energy and by taking into account the $2j+1$ degeneracy 
of each of them. 

In the three magic nuclei considered, $^{12}$C, $^{16}$O  and $^{208}$Pb,
the single particle levels below the Fermi surface are fully occupied.
For these nuclei, the role of the pairing is negligible, meaning that
its effects on binding and single particle energies, and also
on the proton and neutron distributions, are
within the numerical accuracy of our calculations.

%%%%%%%%%&&&&&&&&&&&&%
% BCS occupation
%%%%%%%%%%&&&&&&&&&&
\begin{table}[htb] 
\begin{center} 
\begin{tabular}{ c c ccc c ccc } 
\hline \hline 
&~~~& \multicolumn{3}{c}{protons} &~~~& \multicolumn{3}{c}{neutrons} \\
\cline{3-5}\cline{7-9} 
      && s.p. state & ~~$2j+1$~~ & occupancy && s.p. state & ~~$2j+1$~ & occupancy \\
\hline 
$^{20}$Ne & & $1d_{5/2}$  & 6 & 2 &&  $1d_{5/2}$  & 6 & 2 \\
\hline 
$^{40}$Ar & & $1d_{3/2}$  & 4 & 2 &&  $1f_{7/2}$  & 8 & 2 \\
\hline 
$^{76}$Ge & & $2p_{3/2}$  & 4 & 4 && $1g_{9/2}$  & 10 & 4  \\
 \hline 
$^{132}$Xe & & $2d_{5/2}$  & 6 & 4 && $1h_{11/2}$  & 12 & 8  \\
\hline \hline
\end{tabular}
\caption{\small
The s.p. Fermi levels for the open shell nuclei considered in the present work. The $2j+1$ column indicates the number of nucleons when the levels are fully occupied. In the column labelled as ``occupancy'', we give the number of nucleon occupying the Fermi level for the specific isotope. 
}
\label{tab:occu} 
\end{center} 
\end{table}

In the case of the other nuclei studied, $^{20}$Ne, $^{40}$Ar, $^{76}$Ge and $^{132}$Xe, the s.p. levels
with maximum energy are only partially occupied (see table \ref{tab:occu}) and 
we carried out HF calculations in the so-called {\it equal filling approximation},
that is by considering an average occupancy given by the number of nucleons 
occupying the level divided by the total occupancy $2j+1$.
For these nuclei, the effects of the pairing are not any more negligible
and the BCS calculations \cite{rin80,suh07}, performed
on top of the HF ones, permit to take care of them. 
The RC has been carried out 
by using an effective nucleon-nucleon finite-range 
interaction of Gogny type in its
D1M parameterization \cite{gor09} 
in both steps of our calculations, the HF 
and the BCS, where in the latter case it acts as 
pairing force.

The pairing changes the HF occupation numbers 
and this modifies the proton and neutron density distributions that,  
in our model, are obtained as:
\beq
\rho_i(r) \,= \, \sum_{k_i} v_{k_i}^2 \, (2j_{k_i}+1) \, \rr^2_{k_i}(r)  \, , \,\,\, i\equiv {\rm p,n} \, ,
\label{eq:density}
\eeq
where the index
$k_i\equiv (n_{k_i},l_{k_i},j_{k_i})$ labels the s.p. state.
We have indicated with $v_{k_i}^2$  its occupation probability 
and with $\rr_{k_i}(r)$ the radial part of its wave function.  
The sum on ${k_i}$ is restricted to proton or neutron states 
to calculate the respective densities. 

We show in table~\ref{tab:BCS} the occupation probabilities obtained
in our RC, {\it i.e.}  
HF+BCS with the D1M interaction, 
for those s.p. states where the differences with the corresponding HF
values are larger than 0.01, in absolute value. 
This occurs only for $^{40}$Ar, $^{76}$Ge and $^{132}$Xe. 
In the case of $^{20}$Ne this does not happens even if its proton and 
neutron $1d_{5/2}$ s.p. levels are only
partially occupied (see table \ref{tab:occu}).

%%%%%%%%%&&&&&&&&&&&&%
% BCS occupation
%%%%%%%%%%&&&&&&&&&&
\begin{table}[!hb] 
\begin{center} 
\begin{tabular}{ c c ccc c ccc } 
\hline \hline 
&~~~& \multicolumn{3}{c}{protons} &~~~& \multicolumn{3}{c}{neutrons} \\
\cline{3-5}\cline{7-9} 
      && s.p. state~~ & $v_{\rm BCS}^2$ & $v_{\rm HF}^2$ && s.p. state~~ & $v_{\rm BCS}^2$ & $v_{\rm HF}^2$ \\
\hline 
$^{40}$Ar & & $2s_{1/2}$  & 0.824 & 1.000 &&  $1d_{3/2}$  & 0.988 & 1.000 \\
 && $1d_{3/2}$  & 0.591 & 0.500  && &  &  \\
\hline 
$^{76}$Ge & & $2p_{3/2}$  & 0.465 & 1.000 && $2p_{1/2}$  & 0.982 & 1.000  \\
&& $2p_{1/2}$  & 0.041 & 0.000  && $1f_{5/2}$  & 0.977 & 1.000  \\
 && $1f_{5/2}$  & 0.342 & 0.000  && $1g_{9/2}$  & 0.418 & 0.400  \\ 
 \hline 
$^{132}$Xe & & $2d_{5/2}$  & 0.121 & 0.670 && $3s_{1/2}$  & 0.972 & 1.000  \\
 &&$2d_{3/2}$  & 0.012 & 0.000 && $2d_{3/2}$  & 0.940 & 1.000  \\
&& $1g_{7/2}$  & 0.396 & 0.000  && $1g_{7/2}$  & 0.981 & 1.000 \\
&&  $1h_{11/2}$ & 0.010 & 0.000 && $1h_{11/2}$ & 0.705  & 0.670  \\
\hline \hline
\end{tabular}
\caption{\small
Occupation probabilities obtained with the D1M interaction in HF+BCS,
$v^2_{\rm BCS}$, and HF, $v^2_{\rm HF}$, calculations for those s.p. levels
where the differences are larger than 0.01, in absolute value. 
}
\label{tab:BCS} 
\end{center} 
\end{table} 

The quality of the nuclear part of the RC in describing experimental observables can be
deduced by the results presented in table \ref{tab:bene} where
we compare the binding energies per nucleon, $B/A$, and the 
proton and neutron root mean square (rms) radii, calculated as
\beq
r_i \, = \, \left[ 4\, \pi \,  \int {\rm d}r \, r^4 \, \rho_i^{\rm fold} \right]^{1/2} \, , \,\,\, i\equiv {\rm p,n} \, ,
\label{eq:rms}
\eeq
with the experimental values \cite{bnlw,ang13}.

The results of table \ref{tab:bene} show that the
relative differences with the experimental values of  
both the binding energies and the rms radii, 
are smaller than 6.5\%, in absolute value. 
The good description of the nuclear binding energies
is a test of the validity of the general fit carried out to define the
D1M force \cite{cha98}. 
Descriptions of the experimental data of the same quality are 
obtained by using other parameterizations of the Gogny interaction
such as D1S \cite{ber91}, D1ST2a \cite{gra13} or D1MTd \cite {co18a}.

%%%%%%%%%&&&&&&&&&&&&%
% binding energies HF+BCS HFB
%%%%%%%%%%&&&&&&&&&&
%\begin{landscape}
\begin{table}[!t] 
\begin{center} 
\begin{tabular}{c ccc c ccc c ccc} 
\hline \hline
&& \multicolumn{3}{c}{$B/A$ (MeV)} &~~~& \multicolumn{3}{c}{$r_{\rm p}$  (fm)} 
&~~~& \multicolumn{2}{c}{$r_{\rm n}$  (fm)} \\
\cline{3-5} \cline{7-9} \cline{11-12}
            &~~~& HF+BCS  & ~~HFB~~ & ~~exp~ &~~~& HF+BCS  & ~~HFB~~ & ~~exp~~ &~~~& HF+BCS  & ~~HFB~ \\
\hline
 $^{12}$C        && 7.246  & 7.492  &  7.680 && 2.544 & 2.727 & 2.470  && 2.526 & 2.709 \\
 $^{16}$O        && 7.975  & 8.000 &  7.976 && 2.762 & 2.934 & 2.699 && 2.741 & 2.911\\
 $^{20}$Ne      &  & 7.508  &  7.759&  8.032 && 2.951 & 2.832 & 3.006 && 2.911 & 2.802 \\
 $^{40}$Ar       && 8.455  & 8.599    &  8.595 && 3.412 & 3.538 & 3.427  && 3.480 & 3.609\\
 $^{76}$Ge      &  & 8.559  & 8.648     &  8.705 && 4.037 & 4.153 & 4.081  && 4.137  & 4.276 \\
 $^{132}$Xe    &    & 8.327  & 8.376 &  8.428 && 4.765 & 4.861 & 4.786 && 4.861 & 4.991  \\
 $^{208}$Pb     &   & 7.797  & 7.815  &  7.867 && 5.472 & 5.557 & 5.501  && 5.563 & 5.690  \\
\hline 
\end{tabular}
\caption{\small 
Binding energies per nucleon, and proton and neutron rms radii, 
obtained in the HF+BCS approach with the D1M interaction  
and in HFB calculations performed with the SLy5 Skyrme force. 
The experimental values are 
taken from the compilations of Refs.~\cite{bnlw} and \cite{ang13}.
}
\label{tab:bene} 
\end{center} 
\end{table} 

\begin{figure}[!h]
\centering
\includegraphics[width=6.5cm, angle=0]{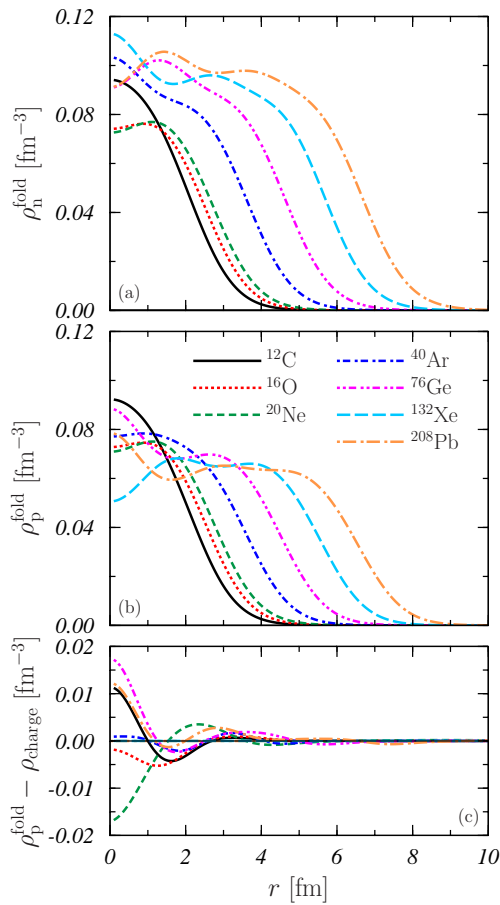} 
\vspace{-0.3cm}
\caption{\small 
Folded neutron (panel (a)) and proton (panel (b)) densities obtained in the  
RC according to eq.~(\ref{eq:rhofold}).
Panel (c) shows the differences with the 
empirical charge distributions taken from the compilation of ref. \cite{dej87}. 
In this last panel, the results for $^{132}$Xe are 
not included because the experimental charge density is not available for this nucleus.
}
\label{fig:dens-diff} 
\end{figure}

The neutron, $\rho_{\rm n}^{\rm fold}$, and proton, $\rho_{\rm p}^{\rm fold}$, folded densities obtained 
in the RC are shown in panels (a) and (b) of Fig. \ref{fig:dens-diff}, respectively.
We compare the $\rho_{\rm p}^{\rm fold}$ densities with the empirical charge distributions 
taken from the compilation of ref. \cite{dej87}. The corresponding differences
are shown in panel (c) where the largest values, of about 0.015 fm$^{-3}$, occur at the nuclear center.

%-----------------------------------------------
% Pointlike and folded densities
%-----------------------------------------------

\begin{figure}[ht]
\centering
\includegraphics[width=6.5cm, angle=0]{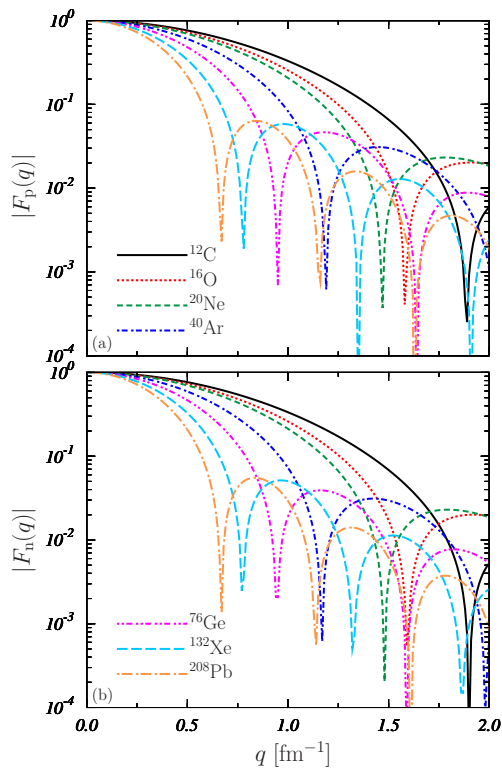} 
\vspace{-0.3cm}
\caption{\small Proton (panel (a)) and neutron (panel (b)) form factors 
as defined in eq.~(\ref{eq:formf-p}), 
obtained in the RC for 
the various nuclei investigated.
}
\label{fig:rcff} 
\end{figure}

The proton and neutron form factors obtained in 
the RC for all the nuclei considered are shown in Fig.~\ref{fig:rcff}. 
The largest is the nucleus the faster is the decrease towards the
first minimum, whose position is located at lower $q$ values.

%------------------------
% RC form factors
%------------------------

\begin{figure}[!h]
\centering
\includegraphics[width=6.5cm, angle=0]{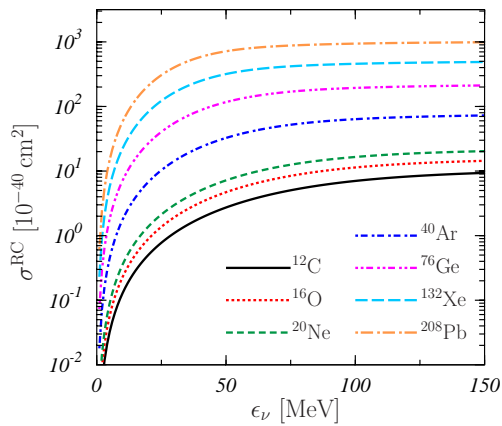} 
\vspace{-0.3cm}
\caption{\small 
Total cross sections obtained in the RC
for all the nuclei considered in the present work. 
%Panel (b) the same cross section, in linear scale, and divided by $N^2$, the neutron number squared. 
}
\label{fig:unoxs-tot} 
\end{figure}

The RC total cross sections for the nuclei considered in the present work
are presented in Fig.~\ref{fig:unoxs-tot}. 
Despite the, expected, quantitative differences, all the curves 
show analogous behavior.
There is a rapid increase of the cross section up to about 30 MeV, 
then this increase slows down and the cross 
sections reach a plateau value after $\sim 50\,$MeV. 
On the other hand the cross 
sections increase with the atomic number, 
those of $^{12}$C\ being two orders of magnitude smaller than those of $^{208}$Pb.

%--------------------------------
% total cross sections
%--------------------------------

%--------------------------------------------------------------------------------------
\subsection{Applications}
\label{sec:super}

We have used the expression (\ref{eq:xstrec})
of the differential cross section to estimate the number of events expected 
for two physical situations.
The first one is related  to the possibility offered by CE$\nu$NS
of detecting neutrinos produced by a supernova explosion
in our galaxy. We have assumed the model of
supernova cooling described in detail in ref. \cite{cos03t}. 
The neutrino emission is characterized by the energy 
distributions, $f_{\nue}$ and $f_{\anue}$ of the 
 electron neutrinos, $\nue$, and antineutrinos, $\anue$, and 
$f_{\nux}$ that takes into account the emission of neutrinos and 
antineutrinos of the other families all together.

The total number of events generated by neutrinos
of a given flavour ($\nue$, $\anue$ or $\nux$) collected by a detector 
situated at a distance $D$ from the source, and
with a detection threshold energy $\etre$ is:
\beq
{\cal N}_\nu (\etre) \,=\, \frac {{\cal I}_{\rm target} \, N_{\nu}}{4\, \pi\, D^2} \,
\displaystyle \int_0^\infty {\rm d}\epsn \,  f^{\rm SN}_\nu(\epsn) \,
\displaystyle \int_{\etre}^{T_{\rm rec}^{\rm max}}  {\rm d}\trec \, 
\frac {{\rm d}\sigma} {{\rm d}\trec} \, ,
\label{eq:nsup}
\eeq
where ${\cal I}_{\rm target}$ indicates  
the number of target nuclei in the detector, 
\beq
N_{\nu} \,= \,\frac {(1 - a_{\rm fl}) \, {\cal E}_B \, s_{\nu}} {\langle \epsilon_{\nu} \rangle}
\eeq
is the total number of neutrinos of flavour $\nu$ emitted by the supernova, and 
$f^{\rm SN}_\nu(\epsilon_{\nu})$ the corresponding energy distribution. 
In the above expression $a_{\rm fl} \simeq 0.01$  indicates the fraction of energy emitted 
in the neutrino de-leptonization burst, ${\cal E}_B$  
is the total energy released by the supernova explosion and $s_{\nu}$ is the fraction 
of these energy carried by each neutrino flavor. Assuming an equal partition 
of the energy between the different
flavours we have $s_{\nue}= s_{\anue} =1/6$ and $s_{\nux} = 4/6$. Finally, 
\beq
\langle \epsilon_{\nu} \rangle \, = \, \int_0^\infty {\rm d} \epsn \, \epsn \, f^{\rm SN}_\nu(\epsn) 
\label{eq:ave-ene}
\eeq
is the average neutrino energy.

In the RC we used a recent energy distribution 
employed in Refs. \cite{tam12,luj14}
and based on a new fit of the
numerical simulations of supernovae explosions \cite{gal17}:
\beq
 f^{\rm SN}_\nu(\epsn) \,=\, {\cal K}\, \frac {\epsn^{\alpha_\nu}\,  \exp \left(- \displaystyle \frac{\epsn}{T_\nu} \right) } 
                     {\Gamma (\alpha_\nu+2) \, T_\nu^{\alpha_\nu+2} } \, .
\label{eq:fII}
\eeq
In the above equation,
$\Gamma(x)$ is the Euler gamma function, and the constant ${\cal K}$ 
has been inserted to normalize the energy distributions to unity. 
The values of the parameters $T_\nu$ and $\alpha_\nu$ for the three neutrino flavors
are given in table \ref{tab:fluencepar}. In this table, we also indicate 
the average energies, for each neutrino flavor, calculated 
with eq.~(\ref{eq:ave-ene}). 

%% %%%%%%%%
% Fluence parameters
%%%%%%%%%%
%\begin{landscape}
\begin{table}[!b] 
\begin{center} 
\begin{tabular}{c c cc c ccc c ccc} 
\hline\hline
&~~~~~& \multicolumn{3}{c}{$f^{\rm  SN}_\nu(\epsn)$} &~~~& \multicolumn{2}{c}{$f^{\rm  MB}_\nu(\epsn)$} &~~~~~& \multicolumn{3}{c}{$f^{\rm  FD}_\nu(\epsn)$}  \\
\cline{3-5}\cline{7-8}\cline{10-12}
&& $T_\nu$ (MeV) & ~~$\alpha_\nu$~~ & $\langle \epsn \rangle$ (MeV)&& $T_\nu$ (MeV) & ~$\langle \epsn \rangle$ (MeV) && $T_\nu$ (MeV) & ~~$\eta_\nu$~ & $\langle \epsn \rangle$ (MeV)  \\ \hline
    $\nue$   & & 2.71  & 2.5 & 9.49 && 3.5 & 10.5 && 2.5 & 4.0  &11.17  \\
    $\anue$ && 3.43  &2.5 & 12.01 && 5.0 & 15.0 && 3.6 & 2.0 &12.98   \\
    $\nux$   && 4.46  & 2.5 & 15.61 && 8.0 & 24.0 && 4.8 & 1.0  & 15.96 \\
\hline \hline
\end{tabular}
\caption{\small Parameters used in the expressions of the 
supernova neutrino energy distributions
(\ref{eq:fII}), (\ref{eq:fB}) and (\ref{eq:fI}) taken from 
Refs. \cite{pap15,gal17}. 
The corresponding average energy values $\langle \epsn \rangle$ are also shown.
%The values of $T_\nu$ and $\langle \epsn \rangle$ are expressed in MeV.
\label{tab:fluencepar} 
}
\end{center} 
\end{table} 
%\end{landscape}

In the RC we considered the neutrinos emitted by a supernova which
releases a  total energy of ${\cal E}_B = 3.0 \times 10^{53}\,$erg 
at a distance $D=10\,$kpc from the 
earth laboratory where a $1\,$t detector is used. 

The second application of our model is related to 
the production of neutrinos induced by a SNS.
In a first step the neutrons generated by the source interact
with a nuclear target and produce  $\pi^+$ pions.
By stopping these pions we obtain monochromatic neutrinos 
from the decay of $\pi^+$ at rest:
\beq
\pi^+ \, \rightarrow \, \mu^+  \,+\,  \num \, .
\label{eq:piond}
\eeq
These are the prompt, monochromatic,  
neutrinos whose energy distribution is
\beq
\fsns_{\num} ( \epsilon_{\num} ) \, = \, \delta( \epsilon_{\num}\, -\, E_{\num})
\eeq
where 
\beq
E_{\num}\, =\, \frac {m^2_\pi \,-\, m^2_\mu} {2\, m_\pi} \,=\, 29.79 \, {\rm MeV} \, .
\eeq
In the previous expression, we have indicated with
$m_{\pi}$ and $m_{\mu}$ the pion and muon mass, respectively.
We neglect the tiny fraction of electron neutrinos 
coming from the $\pi^+$ decay.

In addition to the prompt neutrinos, there are also the delayed neutrinos 
coming from the decay of $\mu^+$:
\beq
\mu^+ \,\rightarrow \, e^+ \,+\, \anum \,+\, \nue \, .
\eeq
The energy distributions of this three-body decay are
well described as \cite{avi03}:
\beqn
\label{eq:nue-SNS}
\fsns_{\nue} ( \epsilon_{\nue} )& = & \frac {96 \, \epsilon^2_{\nue} }{ m_\mu^4} \,
\left( m_\mu \,- \,2\, \epsilon_{\nue} \right)\, \theta\left(\epsilon_{\nue}\,-\,\frac{m_\mu}{2}\right) \, , \\
\label{eq:anum-SNS}
\fsns_{\anum} ( \epsilon_{\anum} ) & = & \frac {16 \, \epsilon^2_{\anum} }{m_\mu^4} \,
\left( 3 \,m_\mu \,-\, 4 \epsilon _{\anum} \right)\, \theta\left(\epsilon_{\anum}\,-\,\frac{m_\mu}{2}\right) \, ,
\eeqn
where $\theta(x)$ is the step, or Heaviside, function.

The total number of events generated by neutrinos 
of a given flavour and collected by a detector with a threshold energy $\etre$, 
is given by
\beq
{\cal N}_\nu (\etre) \,=\, \Phi \, {\cal I}_{\rm target} \, t \, \,
\displaystyle \int_0^\infty {\rm d}\epsn \,  f^{\rm SNS}_\nu(\epsn) \,
\displaystyle \int_{\etre}^{T_{\rm rec}^{\rm max}}  {\rm d}\trec \, 
\frac {{\rm d}\sigma} {{\rm d}\trec} \, ,
\label{eq:nsns}
\eeq
where we have indicated with $\Phi$  the neutrino flux reaching 
the detector and with $t$ the exposure time.

In the case of the SNS, for the RC, 
we have assumed a detection time of one year with a 
detector of $1\,$t situated at a distance from the source such 
as it can receive a flux of $10^7\,$neutrinos per cm$^2$ in one second.

%------------------------
% supernova and SNS neutrinos events
%------------------------
\begin{figure}[!t]
\centering
\includegraphics[width=10cm, angle=0]{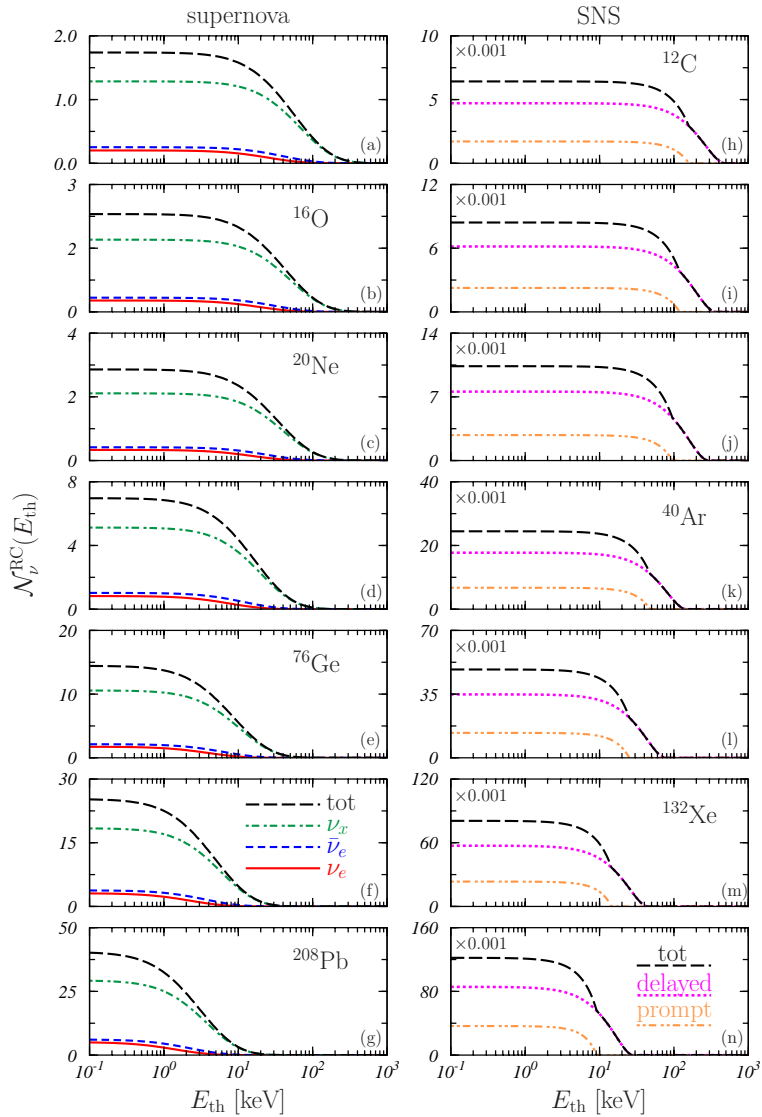} 
\vspace{-0.3cm}
\caption{\small Number of detected events obtained in the RC
as a function of the detection threshold energy
for the cases of supernova explosion (left-handed panels) 
and SNS (right-handed panels), according to Eqs. (\ref{eq:nsup}) and (\ref{eq:nsns}), respectively.
Each couple of aligned panels shows the results obtained for a specific target nucleus.
In the left-handed panels, solid red, dashed blue and  
dashed-dotted green curves 
correspond to $\nu_e$, $\bar{\nu}_e$ and $\nu_x$, respectively. 
In the right-handed panels, the dotted pink and dashed double-dotted orange curves
show the contributions of delayed and prompt neutrinos, respectively.
All the values of the SNS results have been divided by 1000. 
In both supernova and SNS cases, the long-dashed black curves indicate the total number of events.
}
\label{fig:supernu} 
\end{figure}

In Fig. \ref{fig:supernu} the number of events 
obtained in the RC are shown as a function of the detection
threshold energy $E_{\rm th}$.
The results of the left panels (a-g) correspond to the supernova explosion, 
eq.~(\ref{eq:nsup}), while those of the right panels (h-n) have been obtained 
for the SNS, eq.~(\ref{eq:nsns}).  Each couple of aligned 
panels presents results for the same target
nucleus. In each panel, the dashed black curve indicates 
the total number of events, calculated as
\beq
{\cal N}(\etre)\,=\, \sum_\nu {\cal N}_\nu(\etre) \, , 
\label{eq:tot-nev}
\eeq
where the sum runs over the three flavors considered in each case, 
$\nue$, $\anue$ and $\nux$ 
for the supernova, and $\num$, $\nue$ and $\anum$ for the SNS.
In the figure, all the SNS results are divided by a factor 1000. 

As expected, in both cases, the number of events increases
with the mass number of the target nucleus and with the lowering
of the detection threshold energy $E_{\rm th}$. 
At the lowest value of $E_{\rm th}$ we have considered, $10^{-1}$ keV,
the number of the expected events in $^{208}$Pb is 20 times larger
than in $^{12}$C.
On the other hand, light nuclei extend their sensitivity at 
higher values of $E_{\rm th}$. 
The value of the detection threshold energy at which ${\cal N}(\etre) =0$ in $^{12}$C 
is about one order of magnitude larger than in $^{208}$Pb.

In the case of supernova neutrinos, the smallest
contribution is given by $\nue$,  solid red curves, 
while the largest one is due to $\nux$, dashed-dotted green curves. 
The contribution of $\anue$ (short dashed blue curves) is similar to that of $\nue$.
The main reason for this result is that all the neutrino flavors other than electron
neutrinos and antineutrinos are summed up on the $\nux$ contribution and this collective contribution
is weighted 4/6, while that of $\nue$ and $\anue$ count 1/6 each.

For the SNS neutrinos, we separately show the contribution of the 
prompt neutrinos, dashed double-dotted orange curves, and that of the
delayed ones, dotted violet lines, which is about 
two times larger than the previous one. 
This is mainly due to the fact that there are two delayed neutrino flavors, $\nue$ and $\anum$, 
while the prompt neutrinos are only $\num$.

%--------------------------------------------------------------------------------------------------
\section{Comparison with the reference calculation}
\label{sec:results}

As already pointed out in the introduction, the strategy of our investigation
consists in comparing the results of the RC with those obtained
by changing some of the reference inputs. This comparison will be done
by considering mainly the total cross section, $\sigma(\epsn)$, 
and the total number of events,  ${\cal N}(\etre)$. For this reason, we have considered two 
quantities which emphasize the differences between the results of both
calculations.
For the total cross section, given by eq.~(\ref{eq:xstot}), we have 
used the absolute value of the corresponding relative difference:
\beq
\Delta \sigma(\epsn) \, = \, \frac{\left| \sigma(\epsn) \, - \, \sigma^{\rm RC}(\epsn) \right|}{\sigma^{\rm RC}(\epsn)} 
\,.
\label{eq:sdiff}
\eeq
Here $\sigma(\epsn)$ refers to the new calculation. 
For the total number of events, defined in eq.~(\ref{eq:tot-nev}), 
we have calculated:
\beq
\Delta {\cal N}(\etre) \, = \, \frac{\left| {\cal N}(E_{\rm th}) \, - \, {\cal N}^\rc(E_{\rm th}) \right|}
{{\cal N}^\rc(E_{\rm th}=0)}  
\, ,
\label{eq:nevent-diff}
\eeq
a quantity normalized to the maximum total number of 
events obtained in the RC, ${\cal N}^\rc(E_{\rm th}=0)$.

\subsection{The nuclear form factors}
\label{sec:ff}

\begin{figure}[!t]
%\hspace*{0.5cm}
\begin{minipage}[c]{0.45\linewidth}
\includegraphics[width=6.5cm,angle=0]{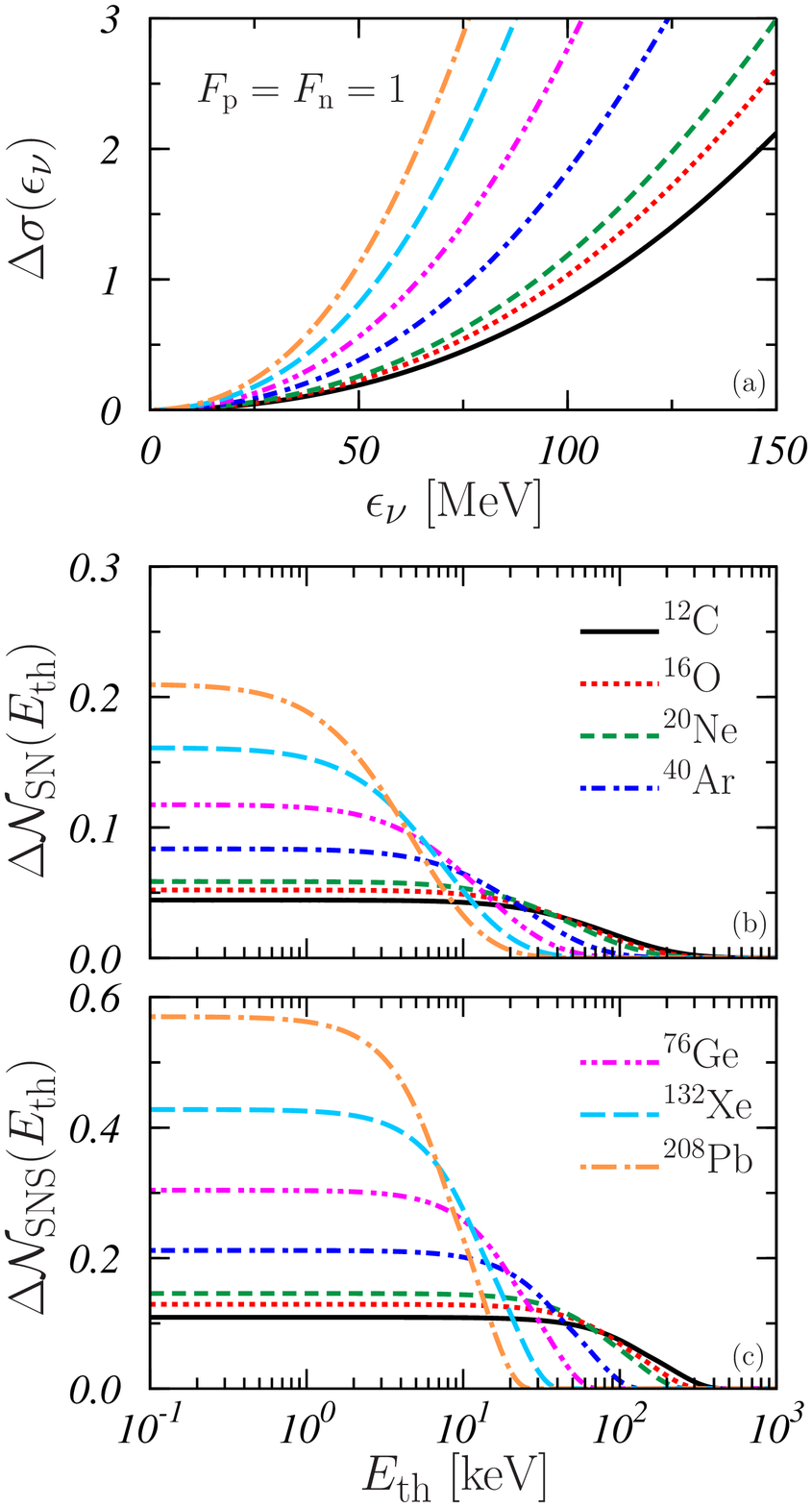}
\vspace{-0.3cm}
\caption{\small 
Relative differences $\Delta \sigma$ (panel (a)), 
defined in eq.~(\ref{eq:sdiff}), and $\Delta {\cal N}$, defined in 
eq.~(\ref{eq:nevent-diff}), for supernova (panel (b)) and SNS (panel (c)), 
obtained by making $F_{\rm p}=F_{\rm n}=1$ in the RC.
}
\label{fig:FpFn1} 
\end{minipage}
\hspace*{0.cm}
\begin{minipage}[c]{0.45\linewidth}
\vspace*{-1.1cm} %\hspace*{-8cm}
\hspace*{-0.2cm}
\includegraphics[width=6.7cm,angle=0]{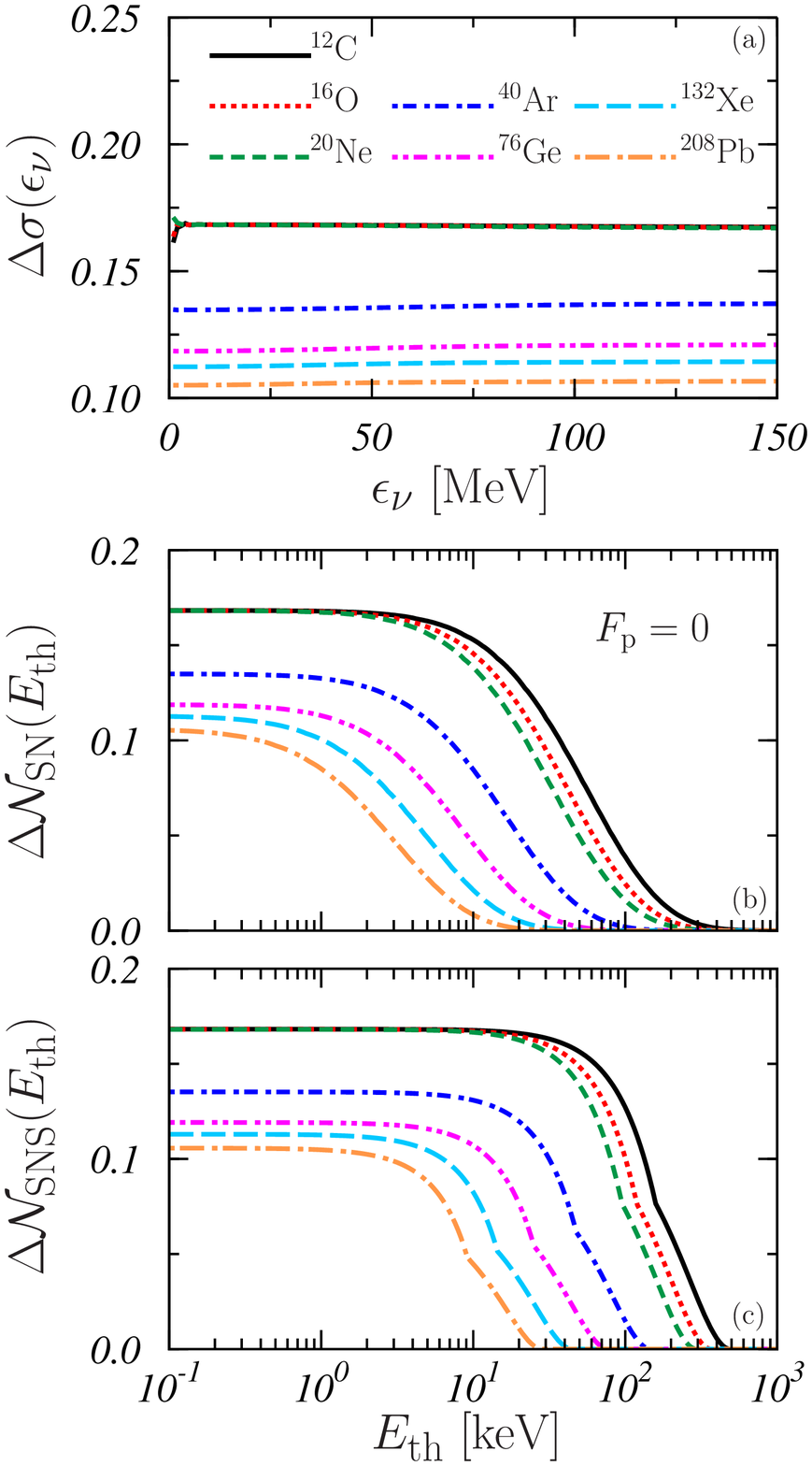}
\vspace{-0.3cm}
%\begin{center}
\caption{\small Same as in Fig.~\ref{fig:FpFn1} but for $F_{\rm p}=0$.
}
\label{fig:Fpzero}
%\end{center} 
\end{minipage}
\end{figure}

If one assumes the low-$q$ limit of the form factors by setting
$F_{\rm p} (q) = F_{\rm n}(q) =1$, the transition amplitude
${\cal A}$, eq.~(\ref{eq:mtrans}),  depends only on the proton 
and neutron numbers, $Z$ and $N$ respectively,
and the total cross section 
(\ref{eq:xstot}) increases as $\epsn ^2$. This can be observed in 
Fig.~\ref{fig:FpFn1}(a) where the relative differences $\Delta \sigma$, 
defined in eq.~(\ref{eq:sdiff}), are shown for all the nuclei studied 
as a function of $\epsn$. As it can be seen, $\Delta \sigma$ 
increases with the nuclear mass.

The effects of this approximation on the 
number of events are shown in
panels (b) and (c) of Fig.~\ref{fig:FpFn1} 
where we present the values of $\Delta {\cal N}$, 
eq.~(\ref{eq:nevent-diff}) for supernova and SNS neutrinos, respectively, as a function
of $\etre$. Despite the quadratic increase of the cross section with
the neutrino energy, the total number of events is limited in both cases.
This is due to the neutrino energy distributions which impose
a maximum value of $\epsn$; 
specifically, about $50\,$MeV in the case of the supernova (see below), and
$m_\mu/2$, in the case of SNS (see Eqs. (\ref{eq:nue-SNS}) and (\ref{eq:anum-SNS})).

%--------------------------------
% Differences Fn=Fp
%--------------------------------
\begin{figure}[ht]
\centering
\includegraphics[width=6.9cm,angle=0]{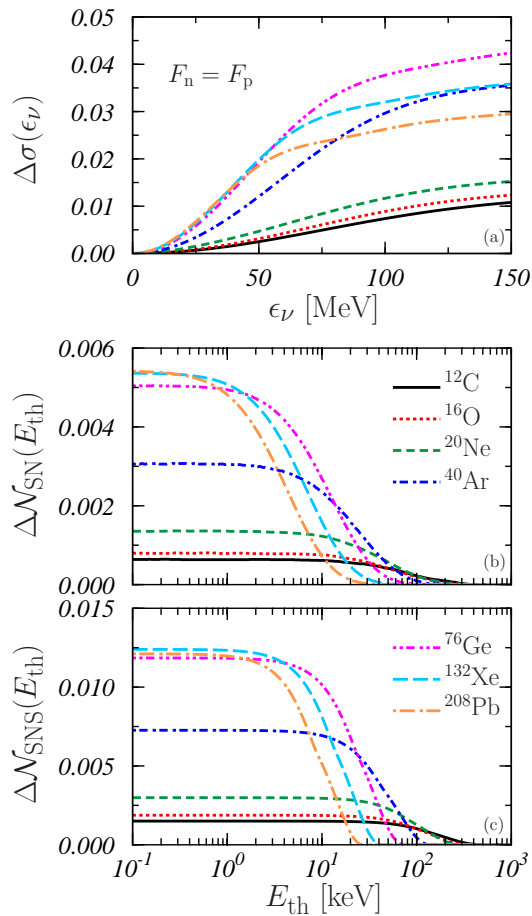}
\vspace{-0.3cm}
\caption{\small Same as in Fig. \ref{fig:FpFn1} but for $F_{\rm n}=F_{\rm p}$.
}
\label{fig:FneqFp} 
\end{figure}

According to eq.~(\ref{eq:mtrans}), the proton contribution to the 
transition amplitude is multiplied by a factor containing the Weinberg angle, 
whose value is about 0.037. This implies that the contribution of 
the protons to the CE$\nu$NS\ cross section is strongly
suppressed, and the process is mainly sensitive to the neutron density
distribution. The relevance of the proton distribution can be deduced 
from Fig.~\ref{fig:Fpzero}, where the quantities $\Delta \sigma$ 
and  $\Delta {\cal N}$, for supernova and SNS, 
calculated by setting $F_{\rm p}=0$, are shown.

As it can be seen in panel (a), the values of $\Delta \sigma$ are 
almost independent of $\epsn$ and coincide for the three nuclei with $N=Z$. 
For the other nuclei we have considered, where $N > Z$, the 
contribution of the protons becomes smaller and, consequently,
the values of $\Delta \sigma$ reduce. As expected, the lowest value 
is that of $^{208}$Pb, the nucleus with the largest difference between
neutron and proton numbers. 

The consequences of neglecting the proton contribution on 
the detected number of events, are shown in panels (b) and (c)
of Fig.~\ref{fig:Fpzero}. The results for supernova and SNS
neutrinos are very similar.
The behavior of $\Delta {\cal N}$ follows that of $\Delta \sigma$. 
For the nuclei with $N=Z$ we obtained similar values of $\Delta {\cal N}$,
about 17\% at $\etre = 0.1\,$keV.
For the other nuclei this value becomes smaller
as $N-Z$ increases until a minimum of $\sim 10\%$ is reached in $^{208}$Pb.

The results of Fig.~\ref{fig:rcff} show that there are not large differences
between the proton and neutron nuclear form factors used in the RC. 
Since the charge distributions are known for a remarkable 
number of nuclei \cite{dej87}, one could use these empirical 
distributions also for neutrons. We have carried out
calculations where we substitute the neutron form factors 
with those of the protons.
The results obtained are compared with those of the RC in 
Fig.~\ref{fig:FneqFp}.

In this case, the values of
$\Delta \sigma$ and $\Delta {\cal N}$ are much
smaller than those presented in the previous two figures. 
The behavior of $\Delta \sigma$ shows an
increase with $\epsn$. This is because the larger is the
neutrino energy, the larger becomes the value of the momentum transfer,
and, therefore, the differences between the form factors are more 
relevant.

The consequences of using the same form factors for protons and neutrons
on ${\cal N}$ are very small 
in comparison with those we have previously discussed.
In general, $\Delta {\cal N}$ grows with $A$ at low $\etre$, and reaches
a value of about 0.5\% for $^{76}$Ge, $^{132}$Xe and 
$^{208}$Pb nuclei in the case of the supernova and roughly 
the double in SNS.

\begin{figure}[!t]
\vspace*{4.5cm}
\begin{minipage}[c]{0.45\linewidth}
\vspace*{0.85cm}
\includegraphics[width=6.5cm,angle=0]{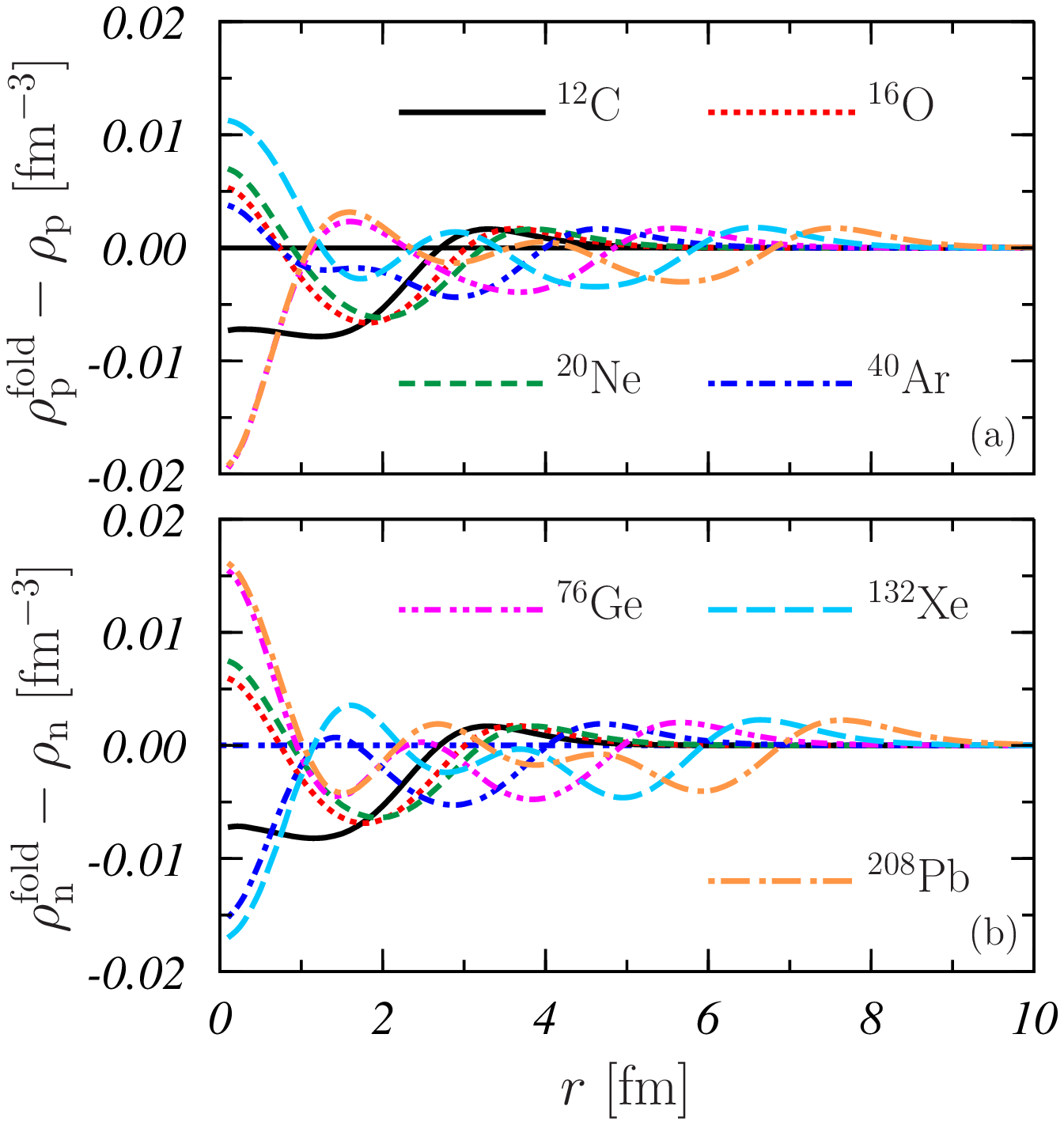}
\vspace{-0.3cm}
\caption{\small 
Differences between folded and point-like density distributions for the nuclei considered
in the present work. Protons (panel (a)) and neutrons (panel (b)) results are shown.
}
\label{fig:dens-fold}  
\end{minipage}
\hspace*{1.cm}
\begin{minipage}[c]{0.45\linewidth}
\vspace*{-5.cm} %\hspace*{-8cm}
\hspace*{-0.2cm}
\includegraphics[width=6.9cm,angle=0]{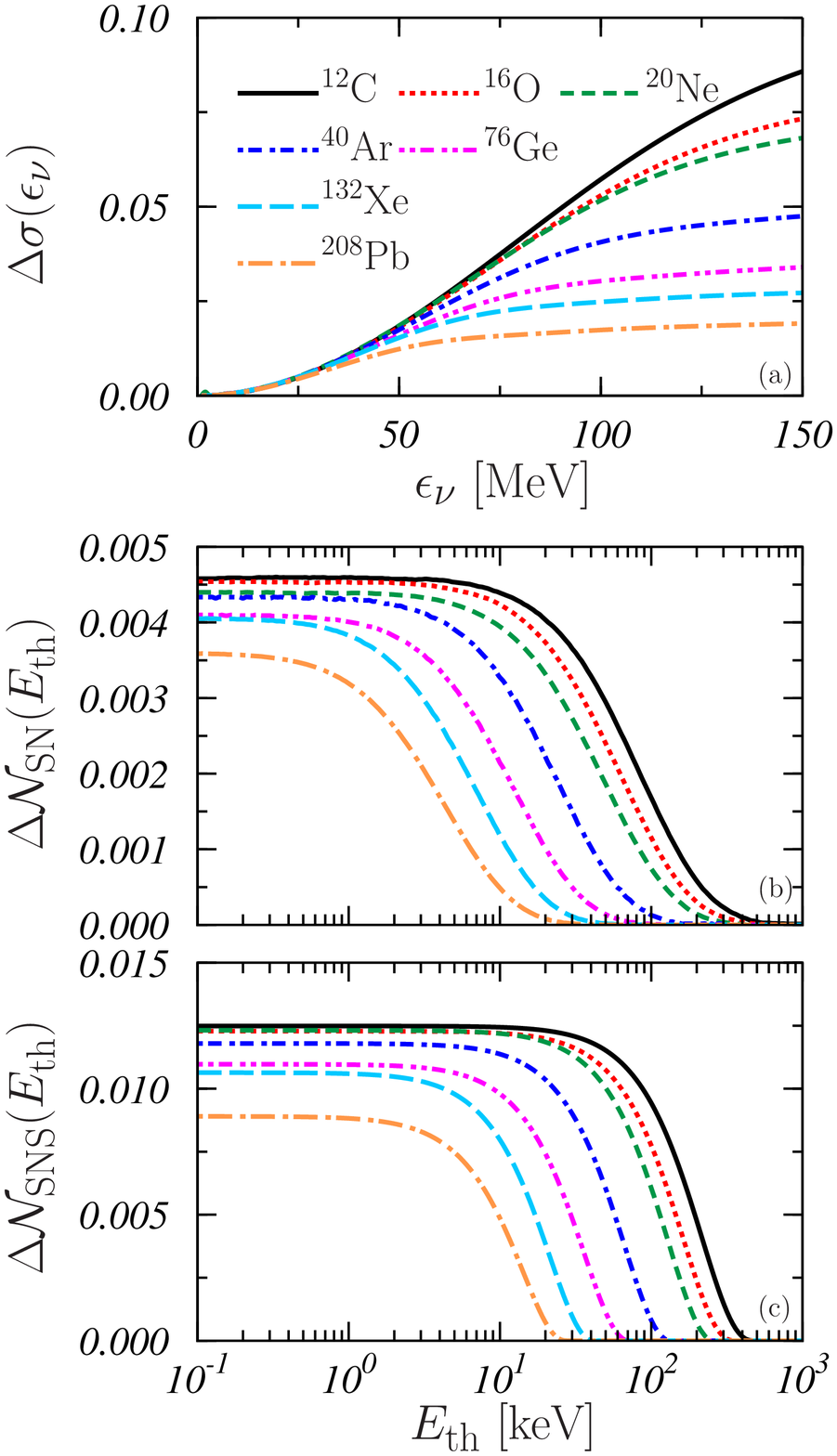}
\vspace{-0.3cm}
%\begin{center}
\caption{\small Relative differences 
$\Delta \sigma$ and $\Delta {\cal N}$ between results obtained
by using point-like and folded densities.
}
\label{fig:fold-point} 
%\end{center} 
\end{minipage}
\end{figure}

In the interaction between neutrino and nucleons we considered
the weak form factor of each nucleon defined in eq.~(\ref{eq:gnge})
to generate the folded nucleon densities according to eq.~(\ref{eq:rhofold}). 
The differences between the point-like distributions 
generated in our HF+BCS nuclear model and those obtained after the
folding with the nucleonic form factor are shown in Fig.~\ref{fig:dens-fold}.
The effects of this folding are more relevant at the nuclear interior where
they reach $\sim 15\%$ of the maximum density values in some of the nuclei studied.
We have also considered
parameterizations of the nucleonic form factor different from that used in our RC (see eq.~(\ref{eq:dipole})), but 
the changes in the folded density distributions are within the numerical accuracy of our calculations. 

%-----------------------------------------------
% Pairing densities
%-----------------------------------------------
\begin{figure}[!b]
\centering
\includegraphics[width=7cm, angle=90]{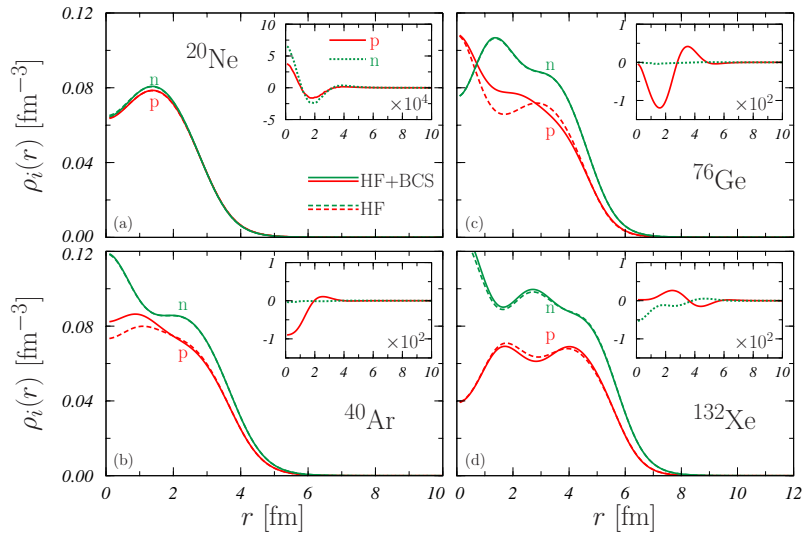} 
\vspace{-0.3cm}
\caption{\small 
Proton (red curves) and neutron (green curves) 
point-like densities obtained according to eq.~(\ref{eq:density}). HF densities (dashed curves)
are compared to the HF+BCS ones (full lines), for those nuclei with s.p. levels
not fully occupied. The insets show the differences between the two distributions.
}
\label{fig:dpair} 
\end{figure}
The differences between the RC results and those obtained by using the point-like nucleon distributions 
are presented in Fig.~\ref{fig:fold-point}. 
The results shown in panel (a) indicate that,  at high $\epsn$,
$\Delta \sigma$ reaches values of about
$10\%$ in $^{12}$C and 2\% in $^{208}$Pb.
We show in panels (b) and (c) the values of
$\Delta {\cal N}$ for supernova and SNS. In both cases 
 these differences are very small, 0.5\% at most in the former case and 1.5\% in the latter one.
 
%--------------------------------
% Differences SC - D1S
%--------------------------------
\begin{figure}[!t]
\centering
\includegraphics[width=7cm,angle=0]{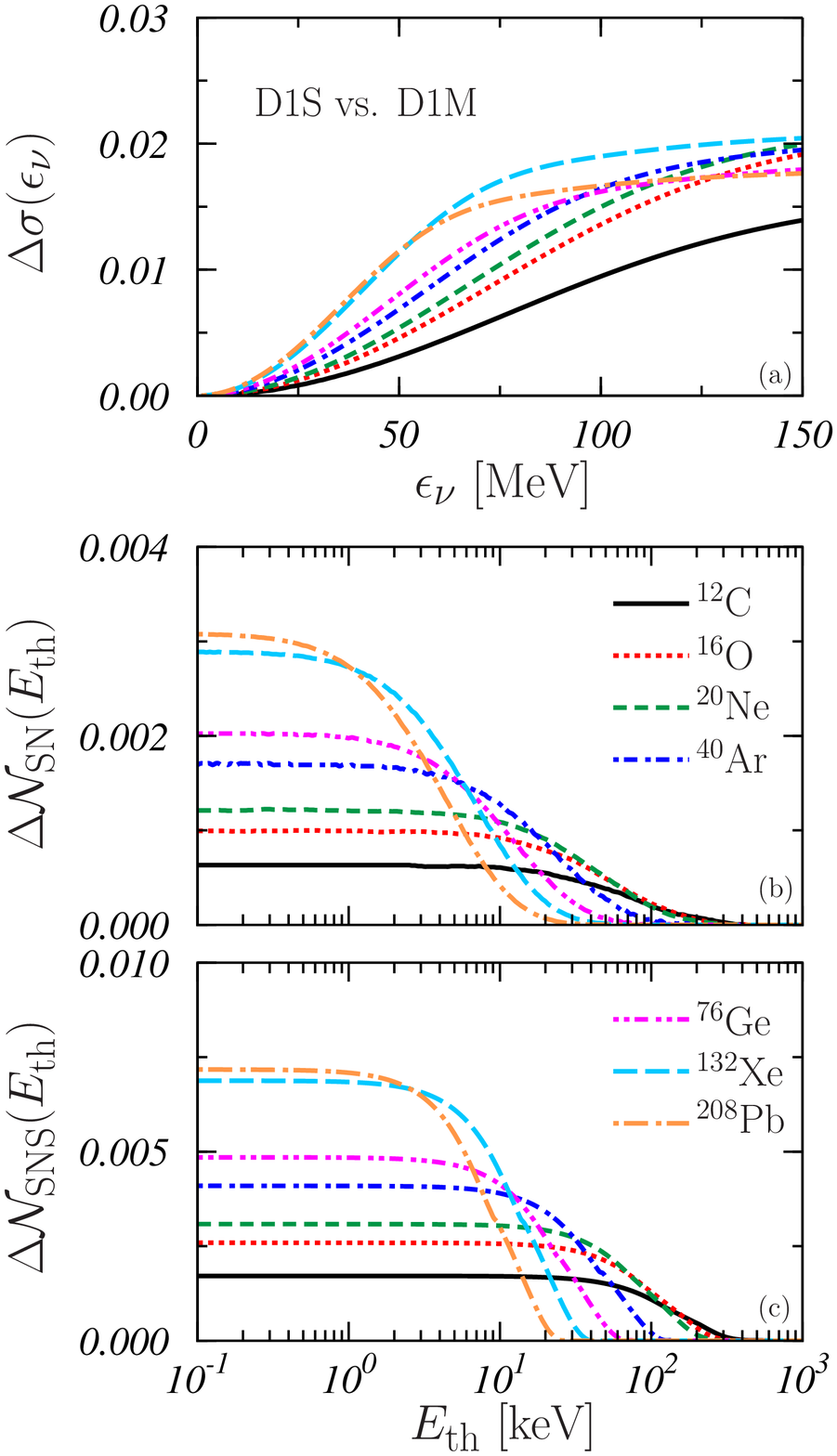}
\vspace{-0.3cm}
\caption{\small Relative differences $\Delta \sigma$ and $\Delta {\cal N}$ 
corresponding to the results obtained by using the D1S interaction instead of the D1M in the RC.
}
\label{fig:diff-D1S} 
\end{figure}

%--------------------------------------------------
\subsection{The nuclear models}

In this section, we study the impact on the quantities of interest 
for CE$\nu$NS processes of the uncertainties related to the various hypotheses
used in developing the nuclear model. 

We consider first the effect of the pairing on the proton and neutron distributions.
In Fig.~\ref{fig:dpair}, the point-like HF densities (dashed lines)
are compared with the point-like densities obtained in a HF+BCS calculation (full lines). 
We have considered here the
four nuclei with open shells where pairing effects are expected 
to be more relevant. 
However, the figure clearly shows that the differences 
between the densities obtained in both calculations
are very small, and their effects on the cross sections and 
on the number of the detected events are within the numerical 
accuracy of our calculations.

Another source of uncertainty in
the nuclear model is related to the effective nucleon-nucleon interaction.
We compare the results of our RC with those obtained 
by carrying out the same HF+BCS calculations
with a different effective
nucleon-nucleon interaction. In these latter calculations we have used  the 
D1S parameterization of the Gogny interaction \cite{ber91} 
instead of the D1M used in our RC. 
The effects on the total cross sections and 
on the number of detected events are shown in Fig. \ref{fig:diff-D1S}. 
The differences in the total cross sections are below 2\% 
while in the total number of events they are smaller than 0.3\%, in 
the case of the supernova, and 0.8\%, in the case of the SNS.

\begin{figure}[!t]
\vspace*{4.5cm}
\begin{minipage}[c]{0.45\linewidth}
\vspace*{0.75cm}
\includegraphics[width=6.5cm,angle=0]{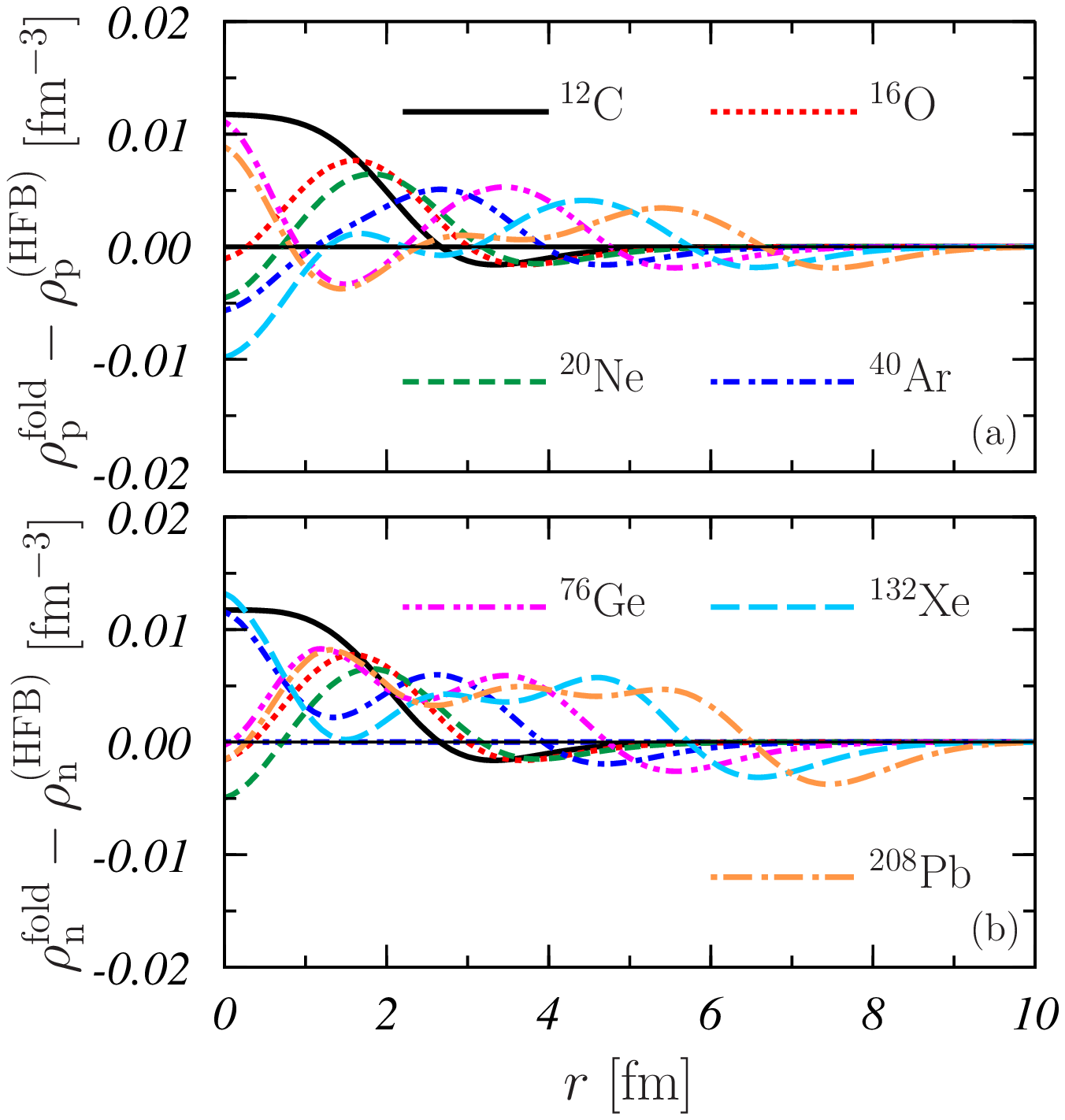}
\vspace{-0.3cm}
\caption{\small 
Differences between the proton, panel (a), and neutron, panel (b),
folded densities used in our RC and those obtained in HFB calculations
with the SLy5 force. 
}
\label{fig:dens-HFB}   
\end{minipage}
\hspace*{1.cm}
\begin{minipage}[c]{0.45\linewidth}
\vspace*{-4.1cm} %\hspace*{-8cm}
\hspace*{-0.2cm}
\includegraphics[width=6.9cm,angle=0]{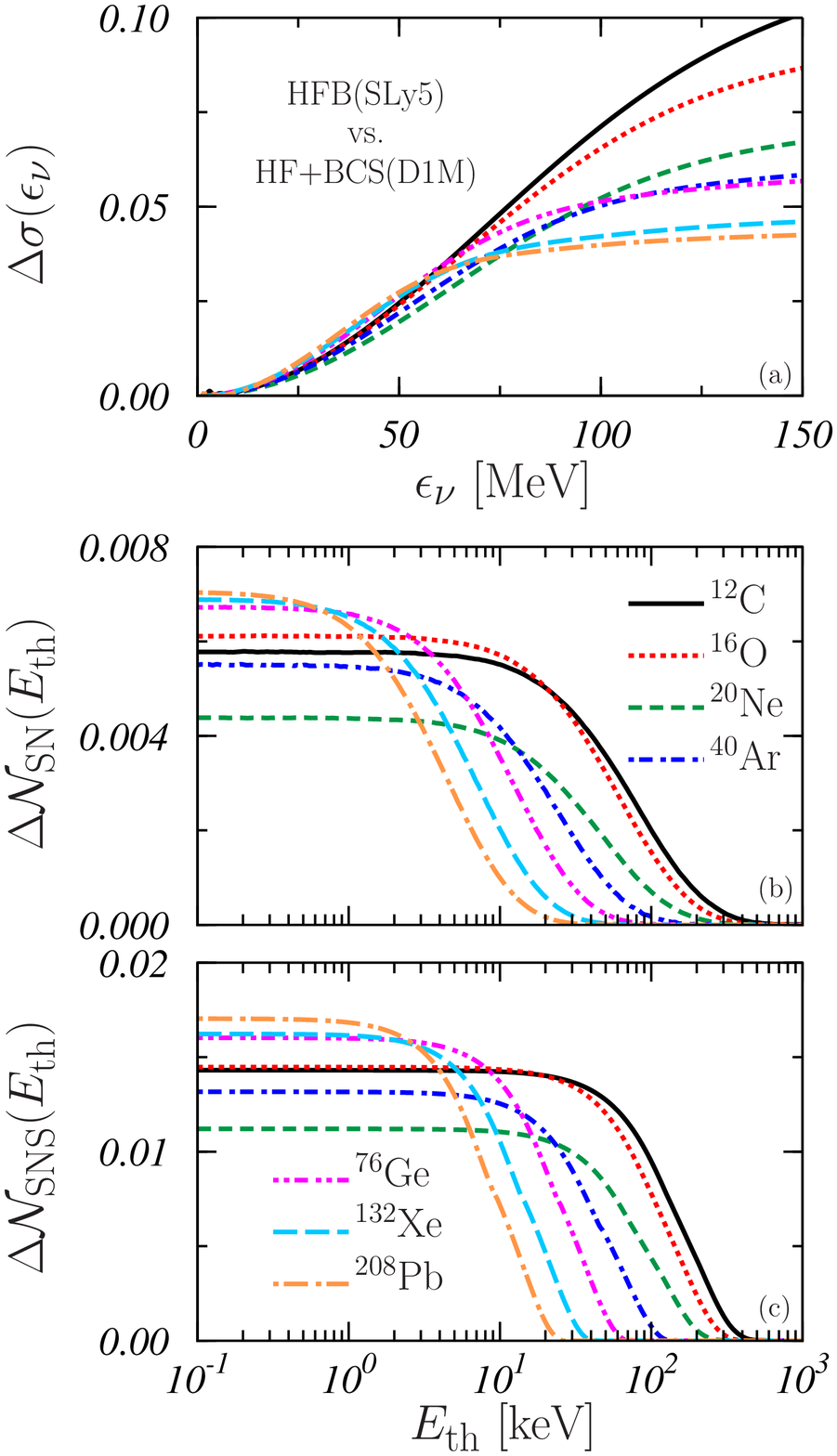}
\vspace{-0.3cm}
%\begin{center}
\caption{\small Relative differences $\Delta \sigma$ and $\Delta {\cal N}$ 
corresponding to the results obtained by using the nucleon distributions obtained in a HFB calculation with the SLy5 Skyrme interaction instead of those considered in the RC.
}
\label{fig:diff-HFB}
%\end{center} 
\end{minipage}
\end{figure}

An additional test of the reliability of our RC 
has been done
by making a comparison with the results of 
Hartree-Fock-Bogolioubov (HFB) 
calculations performed 
with the SLy5 Skyrme interaction.
We have carried out these latter calculations with the HFBRAD
code \cite{ben05}.  As it usually occurs in HFB when a
zero-range effective nucleon-nucleon interaction is used
in the HF sector, the force considered in the pairing sector is not the 
same. In our case, we have used a so-called \textsl{volume} 
pairing field that follows the density shape (see ref. \cite{ben05} for details).

The quality of these HFB calculations in describing the 
empirical values of binding energies and rms charge radii
is indicated in table~\ref{tab:bene}. 
We observe that the performances are rather similar to 
those of our RC calculations. 

We show in Fig.~\ref{fig:dens-HFB} the differences between the folded densities
calculated within our RC and those obtained in the HFB approach.
The largest differences are located at the center
of the nuclei and they are about the 10\% of the maximum 
values of the densities. 

The effects of these differences on the cross sections and on the number of 
detected events are shown in Fig.Ê\ref{fig:diff-HFB}. 
Even though $\Delta \sigma$ reaches values up to about 10\% 
(in the case of $^{12}$C and for $\epsn$=150 MeV), 
the differences $\Delta {\cal N}$ are below 0.8\% and 2\% for the supernova and SNS, respectively. 

Our work is based on the assumption that our model, describing well the experimental values of 
binding energies, charge radii and distributions, provides also 
a good description of the, experimentally unknown, neutron densities.
The recent measurements of the PREX collaboration \cite{PREX12}
indicates, for the $^{208}$Pb nucleus a neutron radius of $r_{\rm n}=5.78^{+0.16}_{-0.18}\,$fm, 
slightly larger than the value of $5.563\,$fm of our RC (see table \ref{tab:bene}). 
Also the estimate of Horowitz {\it et al.} \cite{Hor12},
$r_{\rm n}=5.751\pm 0.175 ({\rm exp}) \pm 0.026 ({\rm model}) \pm 0.005 ({\rm strange})\,$fm,
is larger than our RC result. 
The value of $r_{\rm n}$ measured by the PREX collaboration for $^{208}$Pb is an open
problem, since it is remarkably larger than values obtained with a variety of experimental procedures
which are closer to our RC result \cite{Kra13}.

\begin{figure}[!t]
\centering
\includegraphics[width=7cm,angle=0]{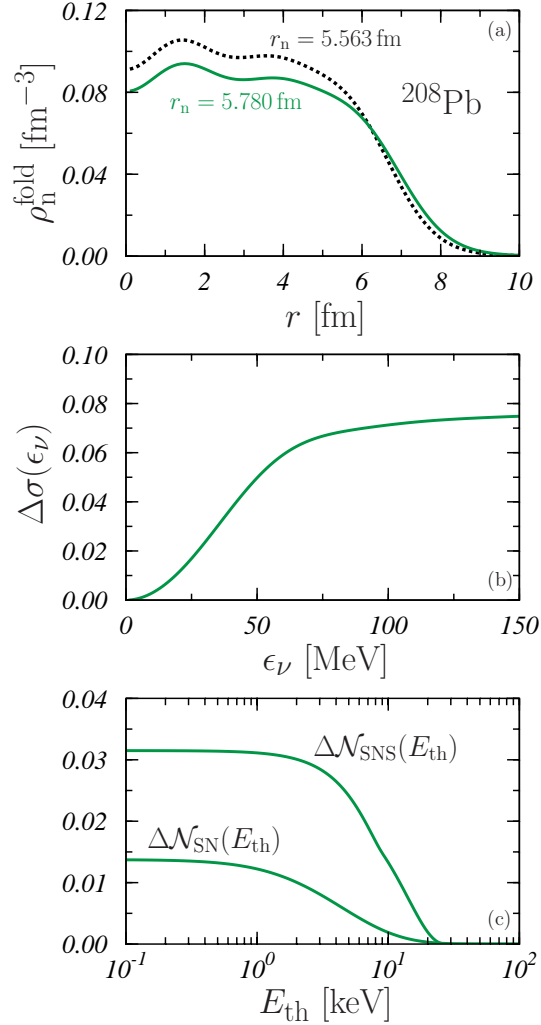}
\vspace{-0.3cm}
\caption{\small Effects of the consideration of an enlarged neutron density for $^{208}$Pb. In panel (a) this density (green solid curve) is compared to that used in the RC (black dotted curves). Panel (b) and (c) show the relative difference differences $\Delta \sigma$ and $\Delta {\cal N}$ obtained for the enlarged neutron density.
}
\label{fig:larger-rho} 
\end{figure}

Despite the fact that the difference between the PREX result and those obtained 
with other techniques has not yet been clarified, we have considered the effects
generated by a neutron density larger than that of our RC. 
For this purpose, we have rescaled our RC neutron density to 
generate a new $\rho^{\rm fold}_{\rm n}$ with $r_{\rm n}=5.780\,$fm. 
In panel (a) of Fig.~\ref{fig:larger-rho} we compare
the new density with that obtained in our RC. 
We calculated the total cross section and the total number of events for supernova and SNS with the rescaled
neutron density. The relative differences with respect to the RC  results 
are shown in panels (b) and (c) of this figure.  

We observe a maximum value of $\Delta \sigma$ of about 8\% at $\epsilon_\nu$ = 150 MeV,
and values of about 3\% for $\Delta {\cal N}$ in the 
SNS case, and even smaller, 1.5\%,  for the supernova
case. These values are of the same order of magnitude of
those obtained by setting $F_{\rm n}=F_{\rm p}$ 
(see Fig.~\ref{fig:FneqFp}) or by using the D1S interaction instead than the
D1M (see Fig.~\ref{fig:diff-D1S}). 
 
%--------------------------------
% Differences SC - D1S
%--------------------------------

\subsection{Uncertainties on neutrino sources}
\label{sec:R2}

The aim of our work is to evaluate the robustness of our RC in 
the evaluation of the CE$\nu$NS cross sections, in order to asses 
its reliability in the investigation of the supernova neutrino sources.
In the previous sections we have analyzed how the uncertainties 
on the various inputs of our nuclear model affect the cross sections and 
the number of detected events in the cases of a supernova explosion and the SNS. 

One of the main goals of the neutrino detection of an eventual
supernova explosion in our galaxy is to identify the temperature and
the average energy related to the energy distribution. This would 
provide important information on the explosion mechanism and 
the cooling phase. Therefore, 
we study now how the number of detected events is sensitive to 
one of the major uncertainties related to the supernova neutrino models in that case:
the energy distribution of the neutrino flux.

In our RC, the neutrino energy distribution considered
for the case of the supernova explosion 
is the $f^{\rm SN}_\nu$ defined in eq.~(\ref{eq:fII}).
There are various plausible alternatives, and among them
we have selected 
a Maxwell-Boltzmann distribution \cite{gio06,pap15},
\beq
f^{\rm MB}_\nu(\epsn) \,=\, \frac{\epsn^2} {2\, T_\nu^3} \, \exp \left( -\, \displaystyle \frac{\epsn}{T_\nu} \right) \, ,
\label{eq:fB}
\eeq
and a Fermi-Dirac like distribution \cite{pag09},
\beq
f^{\rm FD}_\nu(\epsn) \,=\, {\cal C} \,  \frac{\epsn^2} {1\,+\, \exp\left( \displaystyle \frac{\epsn}{T_\nu}\, -\,  \eta_\nu \right) }  \, .
\label{eq:fI}
\eeq
Here ${\cal C}$ is a constant included to normalize to unity the integral of the distribution. The values of $T_\nu$ and $\eta_\nu$ for each neutrino type and the average neutrino energies are given in table~\ref{tab:fluencepar}. 

%-----------------------------------------------
% Supenova fluences
%-----------------------------------------------
\begin{figure}[!t]
\centering
\includegraphics[width=6.5cm, angle=0]{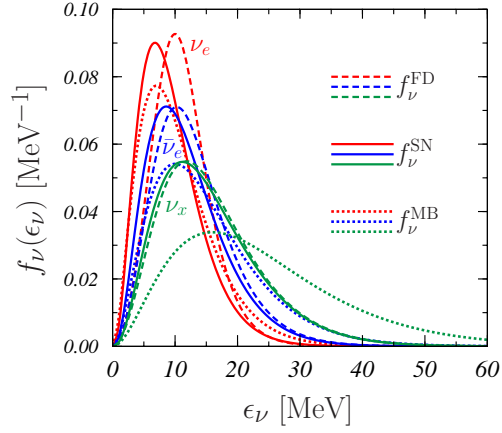} 
\vspace{-0.3cm}
\caption{\small 
Energy distributions of the neutrino emitted by a supernova.
The red curves indicate $\nue$, the blue curves $\anue$ and the green curves $\nux$.
Dotted, dashed and solid curves correspond to 
the $f^{\rm MB}_\nu$, $f^{\rm FD}_\nu$ and $f^{\rm SN}_\nu$ distributions
as given by Eqs. (\ref{eq:fB}), (\ref{eq:fI}) and (\ref{eq:fII}), respectively. 
The parameters shown in table \ref{tab:fluencepar} have been used.
}
\label{fig:fluence} 
\end{figure}

In Fig.~\ref{fig:fluence} we show the behavior of  these energy distributions
for the three neutrino flavors. The most remarkable result is the large difference
between $f_\nu^{\rm MB}$ and the other two distributions, 
in the case of $\nu_x$. In this latter case, the distribution is still
relatively large for neutrino energies above 40 MeV, where all the other
distributions are negligible.

%--------------------------------
% Differences SC - vs. I and B and T/alpha +/-20%
%--------------------------------
\begin{figure}[!b]
\centering
\includegraphics[width=7.5cm,angle=90]{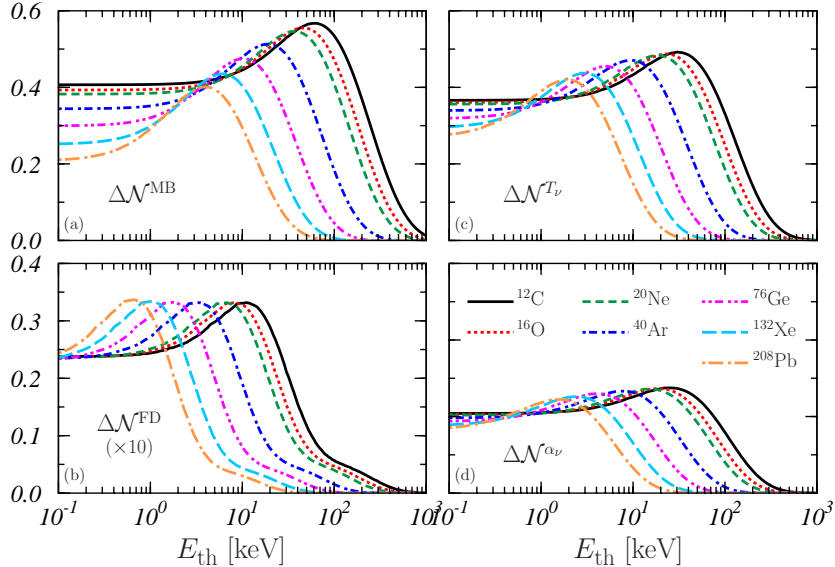}
\vspace{-0.3cm}
\caption{\small Relative differences $\Delta {\cal N}$ between the number of events
from supernova explosion obtained by changing the neutrino energy distribution.
In panel (a) we show the results obtained by using the Boltzmann energy distribution
eq.~(\ref{eq:fB}). In panel (b) those found for the Fermi-like energy distribution
of eq.~(\ref{eq:fI}). The differences shown in panel (c) and (d) are defined in eq.~(\ref{eq:diff-ta}) and correspond to varying by $\pm 20\%$ the values of the parameters $T_\nu$ and $\alpha_\nu$ of $f$ given in table~\ref{tab:fluencepar}. Note that the values of the panel (b) are multiplied by 10.
}
\label{fig:fnueff} 
\end{figure}

In panels (a) and (b) of Fig.~\ref{fig:fnueff} we show the relative differences $\Delta {\cal N}$
between the results obtained by using $f^{\rm MB}_\nu$ and $f^{\rm FD}_\nu$ 
instead of $f^{\rm SN}_\nu$ in the RC. 
These differences are remarkable larger
than those obtained by changing the nuclear physics inputs
(compare with the analogous results shown in Figs.~\ref{fig:diff-D1S} and \ref{fig:diff-HFB}).

The calculations done with $f^{\rm FD}_\nu$ (panel (b)) produce differences $\Delta {\cal N}^{\rm FD}$ 
up to about 3.5\% (the results in the figure are 
multiplied by a factor 10). 
The values of $\Delta {\cal N}^{\rm MB}$ (panel (a)) 
are much larger at the maximum, ranging from about $40\%$ in $^{208}$Pb up to roughly
$60\%$ in $^{12}$C.
We expected such large values because of the 
strong differences between $f^{\rm SN}_\nu$ 
and $f^{\rm MB}_\nu$ observed in Fig.~\ref{fig:fluence}. 

We further estimate the sensitivity of our results to the neutrino
energy distributions by carrying out calculations with different values 
of the $T_\nu$ or $\alpha_\nu$ parameters in the $f_\nu$ energy 
distribution.
In panels (c) and (d) of Fig.~\ref{fig:fnueff} we show
the relative differences
\beq
\Delta {\cal N}^{\lambda}(E_{\rm th}) \, = \, \frac{{\cal N}^{\lambda +}(E_{\rm th}) \, - \, {\cal N}^{\lambda-}(E_{\rm th}) }{{\cal N}^{\rm RC}(E_{\rm th}=0)}\, ,\,\,\, \lambda \equiv {T_\nu,\alpha_\nu} \, ,
\label{eq:diff-ta}
\eeq
where ${\cal N}^{\lambda+}$ and ${\cal N}^{\lambda-}$ indicate 
the total number of events obtained by increasing and reducing 
the parameter $\lambda$ ($T_\nu$ or $\alpha_\nu$) by 20\%, respectively. 

The values of $\Delta {\cal N}^{T_\nu}$ are 
larger than those of $\Delta {\cal N}^{\alpha_\nu}$ and they are of the same order as the differences found when 
$f^{\rm MB}_\nu$ is used instead of $f^{\rm SN}_\nu$ in the RC. 
In the peak $\Delta {\cal N}^{T_\nu}$
varies between $\sim 50\%$, in $^{12}$C, and $\sim 40\%$, in $^{208}$Pb, 
whereas $\Delta {\cal N}^{\alpha_\nu}$ is
below 20\% in all nuclei considered.

%------------------------------------------------
\section{Summary and conclusions}
\label{sec:conclusions}

The aim of our work was to investigate the effects on CE$\nu$NS 
generated by the uncertainties on the description of the nuclei which are 
the target of the neutrinos. The evaluation of the CE$\nu$NS cross section requires 
the knowledge of the proton and neutron density distributions of the 
target nucleus. We defined a 
type of calculation, which we called RC, containing reasonable, and 
up to date, ingredients for a description of the nuclear ground state. 
The obtained CE$\nu$NS cross sections were used to 
study two specific physical situations: the detection of 
neutrinos coming from supernova 
explosions and those from SNS. We evaluated the expected number of 
detected events for these two cases by considering an ideal detector
and some specific characteristics of the incoming neutrino fluxes.
Our working strategy consisted in modifying the various
inputs of the RC and then by comparing the new results
with those of RC.

We first pointed out the importance of the role of the nucleon form factors,
defined as the Fourier transform of the proton and neutron distributions.
If the low-$q$ limit is considered, $F_{\rm p}=F_{\rm n}=1$ and, though the CE$\nu$NS 
cross section rises as $\epsn^2$, the number of the detected events is limited  
since the neutrino energy distributions have a maximum
energy. In this situation the differences with respect to the RC results are rather 
large arriving at about 60\% for SNS neutrino with $^{208}$Pb\,
target.

The main contribution to the CE$\nu$NS is provided by the neutrons, however,
the presence of the protons is not negligible. 
We found that, if $F_{\rm p}$ is neglected, the effects produced are worth
 up to 15\% in $N=Z$ nuclei.
Clearly, for nuclei with neutron excess the effect becomes smaller, even though 
it is always larger than 10\%.

These two basic requirements of the calculation are those producing the
largest effects. We studied the use of the same form factors
for protons and neutrons, the role of the weak nucleonic form factor, that of 
the pairing, that of the effective nucleon-nucleon interaction, and that
of an alternative nuclear structure model. 
In the case of the $^{208}$Pb nucleus,
we have also investigated the effects produced by a
rescaling of the neutron distribution to obtain 
larger rms radius. All these changes, with respect
to our RC, generate effects that are, at least, one order of magnitude smaller 
than those produced in the two cases quoted above. 
We are talking about effects of the order of few percents. 

We compared these tiny effects due to the nuclear structure with those related 
to the uncertainties on the supernova neutrino flux models. Specifically, we concentrated on 
the energy distributions of the neutrino emitted by a supernova. First we adopted
two, reasonable, energy distributions different from to that used in our RC. In a second
step we modify by 20\% the parameters of the energy distributions used in the RC.
These changes of the neutrino energy distributions produce effects of about 10\%,
at least one order of magnitude larger than those induced by the nuclear structure
uncertainties.
Our results agree with those obtained using other simpler, analytical, nucleon densities and/or form factors \cite{pap15,Ari19}.

We conclude that CE$\nu$NS processes can be reliably used to study
the details of the neutrino sources generated by a supernova explosion, since the
nuclear structure uncertainties generate effects orders of magnitude smaller than
those related to the neutrino emission. 

On the other hand, the use of CE$\nu$NS experiments to get information
on the neutron distribution, as it has been recently envisaged
\cite{pat12,cad18,ciu18,Pay19}, requires a knowledge of the neutrino flux 
of few percent.

%%%%%%%%%%%%%%%%%%%%%%%%%%%%%%%%%%%%%%%%%%%%%%%%%%%%%%%%%%%
\newpage \clearpage

\end{document}